\documentclass[preprint,12pt, a4paper]{elsarticle}

\usepackage{amssymb}
\usepackage{hyperref}
\usepackage{xcolor}
\usepackage{color}
\usepackage{soul}
\biboptions{sort&compress}
\journal{SoftwareX}

\begin{document}
\renewcommand{\labelenumii}{\arabic{enumi}.\arabic{enumii}}

\begin{frontmatter}

\title{\texttt{ANT.UI}: An interactive 3D tool for preparing ANT.Gaussian molecular junction geometries}

\author[1]{A. Martinez-Garcia\corref{cor1}}
\ead{andres.martinez@ua.es}
\author[2,3]{J. J. Palacios}
\author[1]{C. Sabater}

\cortext[cor1]{Corresponding author}

\affiliation[1]{organization={Departamento de Física and Instituto Universitario de Materiales de Alicante (IUMA)},
            addressline={Universidad de Alicante, Campus de San Vicente del Raspeig},
            postcode={E-03690},
            city={Alicante},
            country={Spain}}

\affiliation[2]{organization={Departmento de Física de la Materia Condensada, Universidad Autónoma de Madrid},
            postcode={28049},
            city={Madrid},
            country={Spain}}

\affiliation[3]{organization={Condensed Matter Physics Center (IFIMAC) and Instituto Nicolás Cabrera (INC), Universidad Autónoma de Madrid},
            postcode={28049},
            city={Madrid},
            country={Spain}}

\begin{abstract}

\texttt{ANT.UI} is a Python graphical interface that automates the construction of molecular-junction geometries for NEGF-DFT quantum transport calculations. Through a real-time 3D viewer, users interactively position electrodes and molecules and generate complete, ready-to-run input files for \texttt{Gaussian} and \texttt{ANT.Gaussian} without manual scripting. Dedicated Pull, Grid, and Rotation assistants further automate electrode-pulling sequences, surface scans, and step-wise rotation studies, with optional geometry-optimisation chaining across each sequence. By replacing a process that previously demanded days of custom scripting with a point-and-click workflow, \texttt{ANT.UI} accelerates research in theoretical molecular electronics and lowers the barrier to entry for new users. The software also exports all constructed geometries in standard XYZ format, allowing direct reuse in molecular dynamics codes or third-party visualization tools without manual reformatting.
\end{abstract}


\end{frontmatter}

\section*{Metadata}
\label{}

\begin{table}[!h]
\begin{tabular}{|l|p{6.5cm}|p{6.5cm}|}
\hline
\textbf{Nr.} & \textbf{Code metadata description} & \textbf{Metadata} \\
\hline
C1 & Current code version & v1.0.0 \\
\hline
C2 & Permanent link to code/repository used for this code version & \url{https://github.com/And3mg/ANT.UI} \\
\hline
C3  & Permanent link to Reproducible Capsule & Not available \\
\hline
C4 & Legal Code License   & MIT License \\
\hline
C5 & Code versioning system used & git \\
\hline
C6 & Software code languages, tools, and services used & Python 3.8+; libraries: tkinter, customtkinter, numpy, matplotlib \\
\hline
C7 & Compilation requirements, operating environments \& dependencies & Python $\geq$3.8; numpy $\geq$1.21; matplotlib $\geq$3.5; customtkinter $\geq$5.2; tkinter (included with Python on Windows and macOS; system package on Linux). An \texttt{ANT.Gaussian} installation on a high-performance cluster (HPC)  is required to run the generated input files. \\
\hline
C8 & If available, link to developer documentation/manual & Included in the repository as \texttt{ANT.UI\_UserManual.pdf}. \\
\hline
C9 & Support email for questions & andres.martinez@ua.es \\
\hline
\end{tabular}
\caption{Code metadata}
\label{codeMetadata}
\end{table}

\begin{table}[!h]
\begin{tabular}{|l|p{6.5cm}|p{6.5cm}|}
\hline
\textbf{Nr.} & \textbf{(Executable) software metadata description} & \textbf{Please fill in this column} \\
\hline
S1 & Current software version & 1.0.0 \\
\hline
S2 & Permanent link to executables of this version  & \url{https://github.com/And3mg/ANT.UI/blob/master/ANT.UI_v1.0.0.zip} \\
\hline
S3  & Permanent link to Reproducible Capsule & Not available \\
\hline
S4 & Legal Software License & MIT License \\
\hline
S5 & Computing platforms/Operating Systems & Microsoft Windows, Linux, macOS \\
\hline
S6 & Installation requirements \& dependencies & Python $\geq$3.8; install dependencies with \texttt{pip install -r requirements.txt} \\
\hline
S7 & If available, link to user manual & Included in the repository as \texttt{ANT.UI\_UserManual.pdf} \\
\hline
S8 & Support email for questions & andres.martinez@ua.es \\
\hline
\end{tabular}
\caption{Software metadata}
\label{executabelMetadata}
\end{table}

\section{Motivation and significance}

The advancement of molecular electronics relies not only on physical electrical transport experiments but also depends critically on theoretical modelling, encompassing both molecular dynamics simulations and state-of-the-art quantum transport calculations. In this context, the ability to visualise and interact directly with the system's geometry is essential. Precisely understanding how molecules deposit onto a surface or anchor to electrodes is a crucial prerequisite for any rigorous analysis.

To address these challenges, the scientific community relies on highly specialized computational tools:
\begin{itemize}
\item \texttt{Gaussian} an established code in quantum chemistry, enabling the optimisation of molecular structures and the calculation of fundamental electronic properties \cite{GAUSSIAN09}.
\item \texttt{ANT.Gaussian} (ANT) performs transport calculations on the complete junction using Non-Equilibrium Green's Functions combined with Density Functional Theory (NEGF-DFT) \cite{ANTcode, ANT3, ANT_2, ANT_3, ANT_4}.
\end{itemize}

Despite the power of these tools, a significant practical gap remains in the workflow. Preparing three-dimensional atomistic geometries and formatting the specific input files required by each program is a persistent bottleneck. Researchers have traditionally relied on custom \texttt{Python} \cite{python} or \texttt{MATLAB}\cite{MATLAB} scripts to generate junction geometries, a process that is error-prone and time-consuming.

While several general-purpose molecular visualizers and builders are available, such as \texttt{Avogadro} \cite{avogadro}, \texttt{OVITO} \cite{ovito}, and \texttt{ASE GUI} \cite{ase}, none of these provide native support for the specific prerequisites of Non-Equilibrium Green's Function (NEGF) quantum transport calculations, such as the precise definition of Bethe-lattice electrode regions. Commercial packages like \texttt{QuantumATK} \cite{quantumatk} do offer integrated NEGF-DFT transport functionality, but rely on their own proprietary DFT engine and licensing model, and are not designed to interface with the \texttt{Gaussian}/\texttt{ANT.Gaussian} workflow used throughout this work. Moreover, advanced simulation protocols such as step-wise junction stretching (electrode pulling) or two-dimensional surface scans still require highly specialized and automated workflows that none of these platforms provide in a fully integrated Gaussian/ANT-compatible manner.

\texttt{ANT.UI} was developed specifically to bridge this gap for ANT. The user launches the application, selects electrode and molecule geometries from built-in libraries or custom files, positions all components interactively in the real-time 3D viewer, and sets DFT parameters via drop-down menus. The software then generates complete, ready-to-run input files for \texttt{Gaussian} and ANT. By eliminating manual scripting, \texttt{ANT.UI} transforms a process that previously took hours or days into a task completed in seconds, and provides a visual, accessible entry point for students new to theoretical molecular electronics.

\section{Software description}

\subsection{Software architecture}

\texttt{ANT.UI} is implemented in Python and is structured as four source modules:

\begin{itemize}
    \item \textbf{ANT.UI.py} — the main entry point. It defines the GUI classes (\texttt{tkinter\_window}, \texttt{object\_controls}, and the assistant pop-ups), handles user interaction, and coordinates the overall application state.
    \item \textbf{\_DICS.py} — a data dictionary module containing atomic radii, masses, and colours used for rendering; the list of supported exchange-correlation functionals; the mapping from element symbols to electrode parameters; and the parser for the \texttt{USER\_CONF.txt} configuration file.
    \item \textbf{\_OUT.py} — the output generation module. It contains all functions that write Gaussian (\texttt{.gjf}), ANT (\texttt{.ant}), submission script (\texttt{launchANT.sh}), and XYZ files, as well as the logic for sequential-output types (pull, grid, rotation).
    \item \textbf{\_DFT.py} — the DFT file writer. It assembles basis-set blocks, pseudopotentials, optimisation constraints, and the Python connector scripts that manage the cluster-side geometry-update pipeline.
    \item \textbf{\_SOC.py} — an optional module (loaded at runtime; the SOC output buttons are disabled if it is absent) that writes spin-orbit-coupling input files using the basis sets and parameters stored in \texttt{SOC\_params/}.
\end{itemize}

The application relies on \texttt{customtkinter} for the GUI and \texttt{matplotlib} for the embedded 3D viewer. No installation is required beyond the packages listed in \texttt{requirements.txt}.

\subsection{Software functionalities}

\subsubsection*{Interactive 3D junction builder}

\begin{figure}[h]
    \centering
    \includegraphics[width=0.9\linewidth]{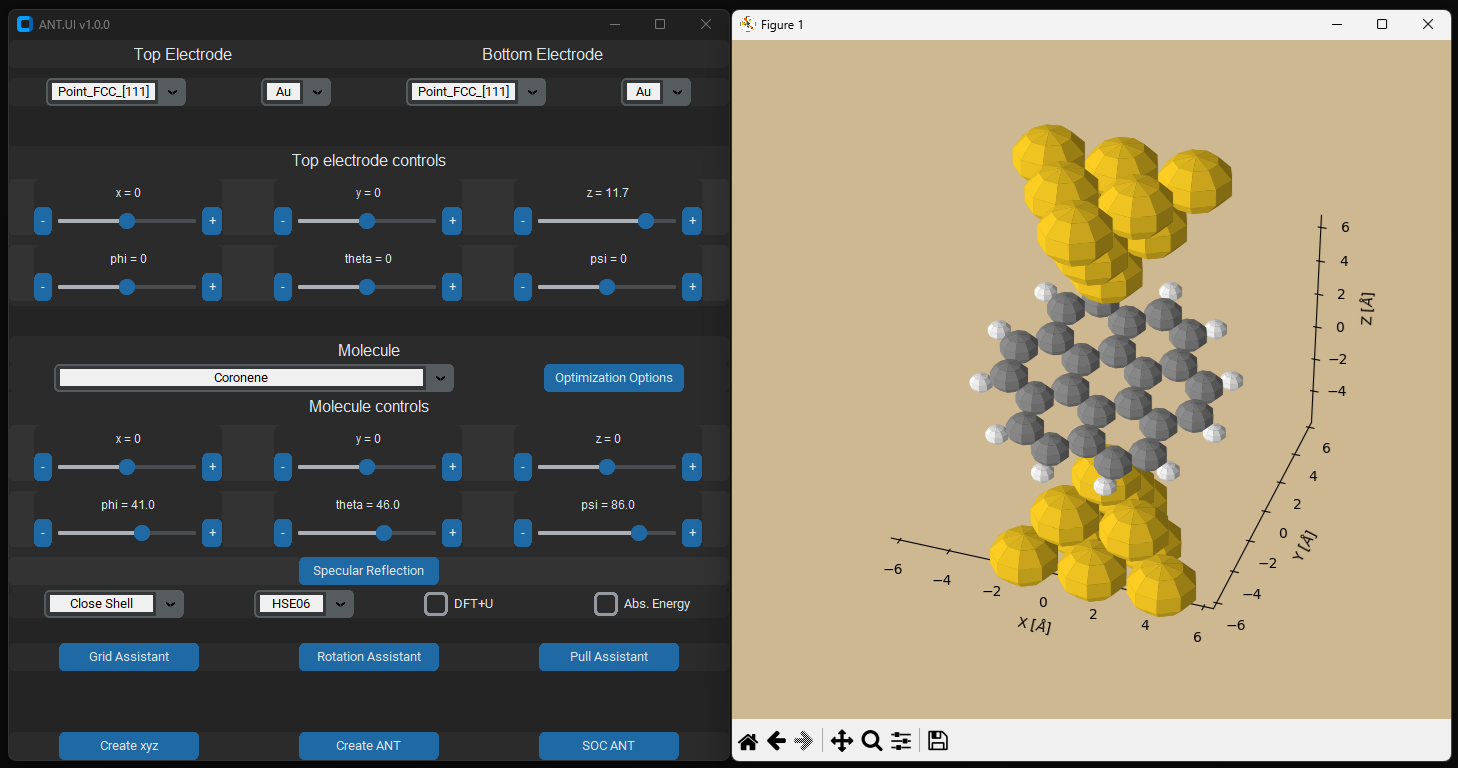}
    \caption{The ANT.UI interface, showing the controller window (left) and the 3D viewer (right).}
    \label{fig:UI}
\end{figure}

The core of \texttt{ANT.UI} is a real-time viewer that renders a 3D representation of the molecular junction using \texttt{matplotlib}'s 3D axes. The user controls the position ($x$, $y$, $z$) and orientation (rotation about each axis) of the top electrode and the molecule via slider controls (see Fig.  ~\ref{fig:UI}), while the bottom electrode acts as a fixed reference frame. Atomic species are colour-coded and rendered as spheres; the Bethe-lattice support layers are hidden from the viewer to avoid visual clutter, since they are treated differently by ANT.

\subsubsection*{Electrode and molecule libraries}

Electrode geometries are stored as XYZ files in the \texttt{ElectrodesDFT} folder. All electrodes are defined with gold as the template element; other elements are obtained by scaling the geometry with the appropriate lattice-parameter ratio from a lookup table in \texttt{\_DICS.py}. The molecule library is likewise XYZ-based, located in the \texttt{Molecules} folder, and is fully extensible: any XYZ file placed in the folder appears in the molecule selector after restarting. DFT+U values for individual functionals can be encoded directly in the XYZ comment line.

\subsubsection*{DFT+U support}

The Hubbard-$U$ correction for strongly correlated molecules is supported at the GGA level. The appropriate $U$ value is pre-computed by the user from isolated-molecule calculations, stored in the molecule's XYZ file, and applied automatically when the DFT+U option is enabled. A user-configurable screening factor (\texttt{U\_ELECTRON\_SCREENING} in \texttt{USER\_CONF.txt}) scales the stored $U$ to account for metallic screening of the molecule by the electrodes.

\subsubsection*{Geometry optimisation integration}

The \textbf{Optimization Options} window provides fine-grained control over which parts of the junction are relaxed by Gaussian and under what constraints. Individual regions (Bethe-lattice layers, active electrode layers, and the molecule) can be independently set to frozen, $z$-only, or fully unconstrained modes. Optimisation can be chained across sequential outputs: the \emph{first step} mode relaxes the initial geometry and propagates it to the rest of the sequence, while the \emph{every step} mode forms a chain where each geometry is initialised from the relaxed previous step.

\subsubsection*{Sequential-output assistants}

Three assistants generate families of related inputs for multi-step calculations:

\begin{itemize}
    \item The \textbf{Pull Assistant} produces a series of inputs in which the electrode separation is varied by a fixed increment $dz$, simulating a break-junction stretching or compression experiment.
    \item The \textbf{Grid Assistant} produces a two-dimensional grid of inputs in which the top electrode is scanned laterally over the molecule at constant height, mimicking an STM measurement.
    \item The \textbf{Rotation Assistant} produces a rotational sweep of a selected junction component about the $z$ axis, enabling systematic studies of orientation-dependent transport.
\end{itemize}

All three assistants respect the optimisation chain settings and generate the necessary Python connector scripts for cluster-side geometry propagation.

\subsubsection*{Spin-orbit coupling}

For electrode elements with parameterised SOC basis sets (Au, Pt, Cu, Ag, Al, Ni), the \textbf{SOC ANT} button generates a parallel set of ANT inputs using the SOC-adapted basis functions and parameters stored in \texttt{SOC\_params/}. The SOC option is also available in the Rotation Assistant for step-wise SOC calculations.

\subsubsection*{User personalization}

ANT.UI creates files for HPC calculations, and many variables will depend on the user, the HPC environment, and the current objectives of the calculation. To accommodate the wide array of desired outputs, we have included the \texttt{USER\_CONF.txt} configuration file. With this file, the user can adapt ANT.UI to their HPC setup and use its outputs without modifying each generated file individually. Users are advised to review this file before using ANT.UI; the user manual in the repository provides a detailed guide to its configuration.

\section{Illustrative examples}

\begin{figure}[htpb]
    \centering
    \includegraphics[width=0.99\linewidth]{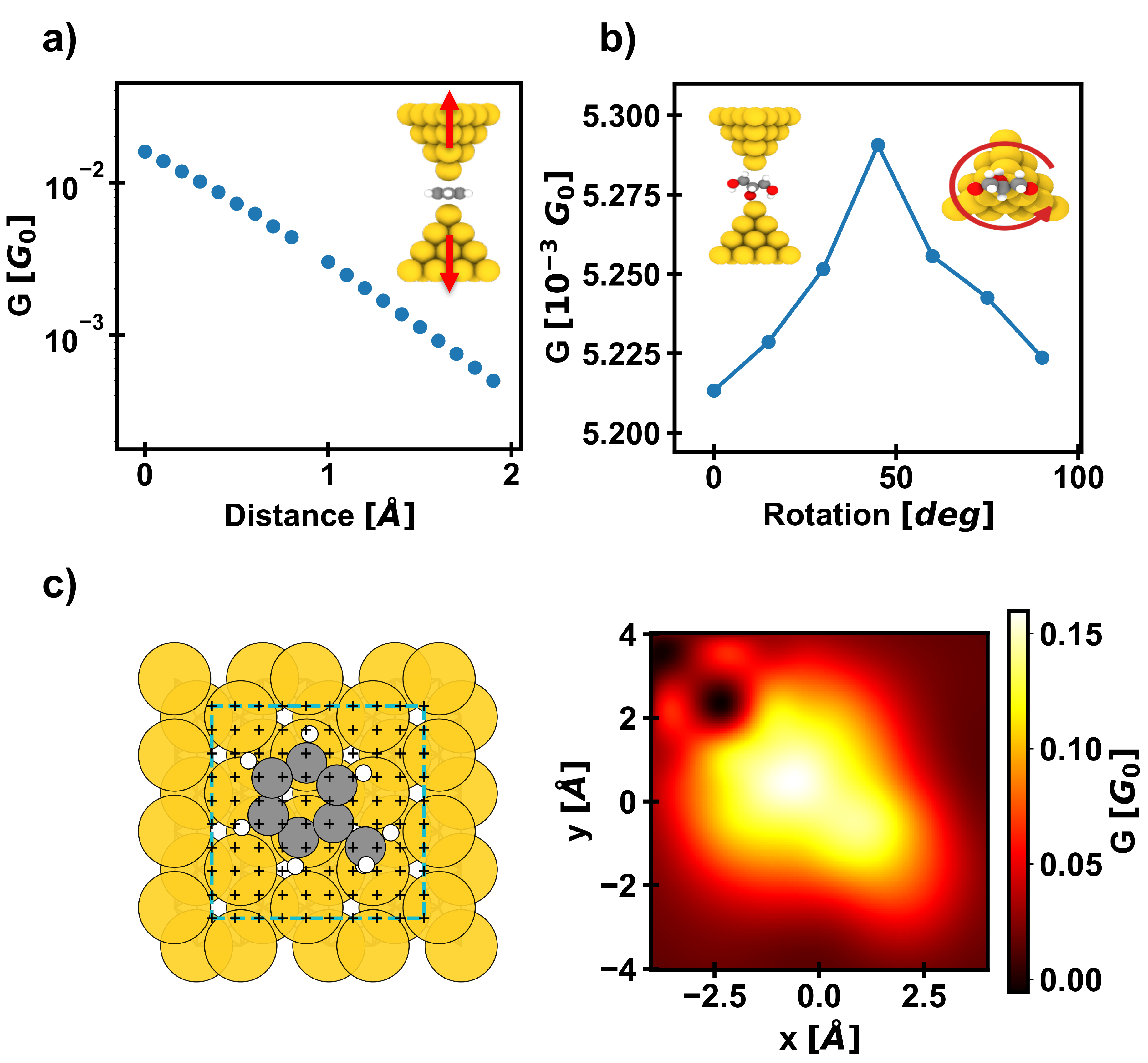}
    \caption{Examples of sequential-output calculations enabled by \texttt{ANT.UI}, each using first-step-only geometry optimization. (a)~Transmission vs.\ electrode separation from a symmetric pulling sequence on benzene. (b)~Transmission vs.\ rotation angle of a glycerol molecule relative to the electrodes. (c)~Grid scan of a toluene molecule on an Au(111) slab. Left: top view of the slab and molecule, with a cross marking each scan point. Right: 2D contour map interpolated from the data.}
    \label{fig:examples}
\end{figure}

Figure~\ref{fig:examples} presents three representative calculations made in ANT and Gaussian from input files generated with \texttt{ANT.UI}, demonstrating the interface's capacity to automate scanning sequences under first-step-only geometry optimization. First, panel (a) shows the zero-bias transmission of a benzene junction as a function of electrode separation, configured via the Pull Assistant. The observed exponential decay with distance accurately captures the expected quantum tunneling behavior. Second, panel (b) illustrates the orientation-dependent transmission of a glycerol molecule rotated relative to the electrodes, seamlessly parameterized by the Rotation Assistant. Finally, panel (c) presents a Grid Assistant scan mapping the transmission of a toluene molecule physisorbed on an Au(111) slab at a constant tip height across a $10\times10$ grid. The left-hand view shows the slab and molecule with a cross marking each scan point, while the right-hand panel shows the resulting 2D contour map interpolated from the data. The spatial variation in zero-bias transmission reveals the nodal structure of the molecular orbitals contributing to transport.

\section{Impact}

The impact of \texttt{ANT.UI} on the computational chemistry and materials physics communities is substantial, directly addressing the following key areas:

\begin{itemize}
    \item \textbf{Enabling new research questions:} By eliminating the bottleneck of manual geometry preparation, \texttt{ANT.UI} makes high-throughput screening of molecular junctions practical. Studies that previously required weeks of input preparation, such as systematic electrode-pulling or rotation sequences across families of molecules, can now be set up in minutes.

    \item \textbf{Enhancing existing research workflows:} The tool improves the reproducibility and accuracy of simulations. Manually defining electrode regions or Bethe-lattice layers in a text file is error-prone and can waste valuable supercomputing allocations. Visual selection and automated exporting in \texttt{ANT.UI} ensure that the geometry being computed corresponds exactly to the researcher's intended design.

    \item \textbf{Transforming daily user practices:} Researchers can focus on analysing physics rather than managing coordinate formats. The software also democratises access to NEGF-DFT calculations by providing an accessible entry point for undergraduate and early-stage doctoral students who do not yet have advanced scripting skills.
    
     \item \textbf{Proven utility in high-impact research:} The reliability and efficiency of {\texttt{ANT.UI}} have been practically demonstrated in peer-reviewed literature. Prior to its formal release, the software's core workflow was successfully employed by our group to generate the complex atomistic input geometries required to study off-resonant electronic transport mechanisms in dithia[$n$]helicene molecular junctions \cite{deAra2024}. This underscores the code's capability to handle advanced experimental and theoretical cross-validations.

\item \textbf{Adoption and reach:} \texttt{ANT.UI} is the standard tool for all members of the Quantum Transport Group at the Universidad de Alicante and of the group AtomEliX at the Universidad Autónoma de Madrid. Moreover, Dr.\ Linda Zotti (Universidad Autónoma de Madrid), Dr.\ Wynand Dednam (University of South Africa), and Dr.\ M.\ P.\ Anantram (University of Washington) are also using this software.
  
\end{itemize}

\section{Conclusions}

\texttt{ANT.UI} provides an integrated, graphical solution to one of the main practical bottlenecks in theoretical molecular electronics: the construction and formatting of molecular junction inputs for NEGF-DFT transport calculations. The software combines an interactive 3D geometry builder with automated output generation for \texttt{Gaussian} and \texttt{ANT.Gaussian}, and extends these capabilities with sequential-output assistants for electrode pulling, surface scanning, and rotational sweeps. It supports geometry optimisation chaining, DFT+U corrections, and spin-orbit coupling inputs for a range of metal electrode elements. By reducing a task that previously demanded extensive manual scripting to a point-and-click workflow, \texttt{ANT.UI} accelerates research productivity and lowers the barrier to entry for new practitioners in the field.
Furthermore, while its primary focus is input generation for quantum transport calculations, \texttt{ANT.UI} inherently produces standard XYZ coordinate files of all constructed geometries. This feature seamlessly bridges the gap with external rendering tools, allowing researchers to visualize complex molecular electronics scenarios and easily produce high-quality, publication-ready graphics, such as Table of Contents (TOC) figures and graphical abstracts.

\section*{Acknowledgements}
The authors C.S. and A.M.-G. gratefully acknowledge financial support from the Generalitat Valenciana (grant CIDEXG/2022/45), as well as additional funding provided by MICIU/AEI/10.13039/501100011033 and the European Regional Development Fund (ERDF/EU) under project PID2023-1466600B-100. J.J.P. also acknowledges MICIU/AEI/10.13039/501100011033 under projects PID2022-141712NB-C21, PID2025-174732NB-I00, and PCI2026-177428-1;  the María de Maeztu Programme for Units of Excellence in R\&D under project CEX2023-001316-M; the Comunidad de Madrid under project “Disruptive 2D materials” (MAD2D-CM-UAM7); and the European Union M.ERA-NET Joint Call 2025 under ALTMAG project. We also wish to express our sincere gratitude to Dr.\ Linda A.\ Zotti (Universidad Autónoma de Madrid) and to Dr.\ W.\ Dednam (University of South Africa) for their valuable discussions and feedback.

\bibliographystyle{elsarticle-num}
\bibliography{mibibSX.bib}

\end{document}